# In-situ Silicon Doped hBN by High-Temperature Molecular Beam Epitaxy Enables Single Photon Emission

Jiyun Kim[1, 2*], Nika Teran[1, 2], Juliette Plo[3], Jonathan Bradford[4], Guillaume Cassabois[3,5], Amy F. M. Collins[4], Tin S. Cheng[4], Christopher J. Mellor[4], Shery L. Y. Chang[6], Sergei V. Novikov[4], Igor Aharonovich[1, 2*]

[1]School of Mathematical and Physical Sciences, University of Technology Sydney, Ultimo, New South Wales 2007, Australia

[2]ARC center of Excellence for Transformative Meta-Optical Systems, University of Technology Sydney, Ultimo, New South Wales 2007, Australia

[3]Laboratoire Charles Coulomb, CNRS-Université de Montpellier, 34095 Montpellier, France

[4]School of Physics and Astronomy, University of Nottingham, Nottingham NG7 2RD, United Kingdom

[5]Institut Universitaire de France, 75231 Paris, France

[6]School of Materials Science and Engineering and Electron Microscope Unit, Mark Wainwright Analytical Centre, University of New South Wales, NSW 2052, Australia

*Corresponding author: chloe.kim@uts.edu.au, igor.aharonovich@uts.edu.au

*Hexagonal boron nitride (hBN) has emerged as a leading host for optically active quantum defects. Yet introduction of specific impurity species other than carbon remains unexplored. Here, we demonstrate an in-situ silicon doping of hBN grown by high-temperature molecular beam epitaxy (HT-MBE). By systematically varying the growth temperature from 900 to 1390 °C under a constant silicon flux, we establish an optimal window for Si incorporation to host a diverse range of emitters from 430-750 nm at room temperature. By transferring silicon-doped hBN film on $SiO_2$ substrate, we verified that single photon emitter activity was sustained in the hBN, demonstrating compatibility with device integration. The plausible origins of the observed optical transitions were discussed, and several potential candidates were proposed. Our results demonstrate a step toward a comprehensive understanding of in-situ doping of hBN and its utilisation for quantum photonic applications.*

Defects in wide band gap semiconductors have been at the centre of attention due to their potential implementation in scalable quantum photonic networks. Among the various hosts, hexagonal boron nitride (hBN) has attracted a sustained interest. It is one of the only few systems that can maintain its wide band gap in both 3D and 2D/few - layered geometries, while also hosting a broad range of luminescent defects. These point defects can be optically isolated and act as ultra-bright quantum emitters, spanning the visible to near-infrared range.

However, despite significant experimental and theoretical efforts, the microscopic origin of many observed quantum emitters remains elusive.[1-3] A variety of intrinsic (vacancies), extrinsic (impurities) as well as complexes and donor acceptor pairs (DAP) have been proposed.[4,5] Nevertheless, the identification of specific defect configurations with tailored emission lines, particularly at the visible spectral range, remains beyond reach.

One of the most studied impurities to date is carbon, given the closely matched atomistic size to the BN host. hBN can be efficiently doped with carbon during the high pressure and temperature growth, which results in yellow coloration of the hBN crystal. Indeed, carbon is believed to be responsible for the common 4.1 eV emission in hBN, commonly observed in these crystals, as well as the B-centres emitting at 436 nm (2.84 eV).[6,7] Furthermore, there is mounting theoretical and experimental evidence that carbon plays a key role at the formation of visible emitters.[8,9] Indeed, various growth techniques successfully doped hBN with carbon and directly correlated the addition of carbon to the presence of bright quantum emitters.[10-12]

In this work, we report the observation and characterisation of silicon (Si) doped hBN. Inspired by the successful studies of group IV defects in diamond (i.e. presence of similar emission patterns from silicon, germanium and tin vacancies complexes in diamond), we develop a robust growth process of hBN doping with Si. Using a Si source in a high-temperature molecular beam epitaxy (HT-MBE) process, we grow Si containing hBN films that host localised quantum emitters. We identify clear photoluminescence (PL) lines from the hBN films that were grown in the presence of Si flux. Our results provide additional opportunities to achieve doping in hBN and expand the range of possible defect configurations. The results further aid towards a more systematic understanding of defect formation and optical activity of quantum emitters in hBN.

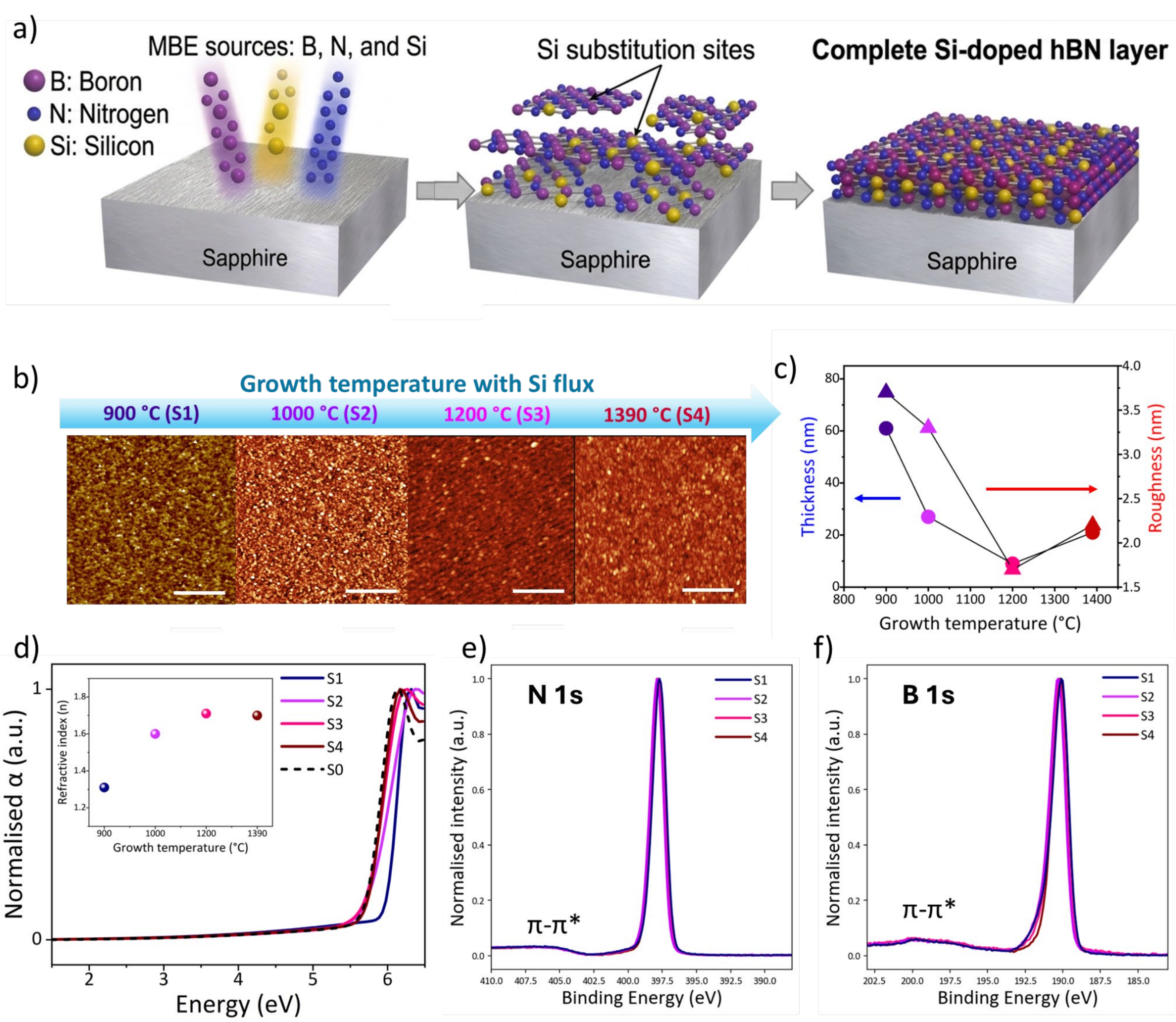

*Figure 1. In-situ Si doping in HT-MBE growth of hBN and characterization of as-grown Si-doped hBN. a) Schematic of MBE growth process with* simultaneous *Si, B, and N beam fluxes, resulting in Si substitutional sites within the hBN lattice. b) AFM images of Si-doped hBN films grown at increasing temperatures: 900 °C (S1), 1000 °C (S2), 1200 °C (S3), and 1390 °C (S4) with a scale bar of 500 nm.*

*c) Thickness and roughness profile as a function of growth temperature (S1-S4). d) Ellipsometry of as-grown hBN films (S1-S4) showing absorption coefficient with an inset of refractive index (n) corresponding to S1-S4. e) and f) XPS spectra of S1-S4, representing N 1s and B 1s binding energies.*

Figure 1 demonstrates an overview of the HT-MBE growth strategy with in-situ Si incorporation and characterisation of the Si-doped hBN samples grown on (0001) sapphire substrates. In our HT-MBE system where a Si sublimation source is installed, we simultaneously supply fluxes of Si, boron and active nitrogen to the growth surface (as shown in Figure 1a), enabling in-situ Si doping at BN lattice sites. To study the mechanisms of Si incorporation during the HT-MBE growth, hBN films were grown across a wide growth temperature range (900-1390 °C) with a fixed Si beam flux. The as-grown Si-doped hBN films are labelled from S1 to S4, corresponding to growth temperature. An pristine hBN film was grown at 1390 °C without Si doping (labelled as S0) as a reference sample for further comparison.

First, the topography of S1-S4 was examined by atomic force microscopy (AFM) to observe surface morphology and roughness. In Figure 1b, the BN films across S1-S4 cover the entire sapphire surface with no significant change in morphology. The roughness of all BN films (S1-S4) is slightly higher than typical MBE grown BN film (< 1 nm), which could be attributed to the Si substitutes causing lattice distortion within the plane. According to the thickness and roughness profile plotted in Figure 1c, the roughness and thickness of the as-grown hBN films decrease with increasing growth temperature. Such trends imply that the film growth rate and quality are governed by the kinetics of adatom (B, N, and Si) incorporation at a given growth temperatures.

Next, we performed spectroscopic ellipsometry to confirm the structural quality of the hBN samples S1-S4. In Figure 1d, all hBN films show absorption near 6 eV, confirming the hBN bandedge. In addition, S3 and S4 demonstrate similar bandedge with the refractive index of ~1.7 in visible range compared to the pristine hBN, S0, whereas S1 and S2 show optical loss and blue-shift in absorption edge with slightly lower refractive index. X-ray photoelectron spectroscopy (XPS) was carried out to quantify the Si sublimation in all samples. As shown in Figure 1e and f, we observe prominent N 1s and B 1s signals at 398.1 eV and 190.5 eV, respectively, which is on par with previous reports.[13] Close inspection of the high-resolution N 1s and B 1s core level spectra reveals π-π* shake-up features indicative of $sp^2$ hBN. The B 1s core level shows slight asymmetry to the high binding energy side which could be indicative of excess boron at lower temperatures that has oxidised upon air exposure. Quantitative analysis of the spectra reveals hBN stoichiometries close to 1:1 regardless of the growth temperature and Si flux. Note that 102 eV peak corresponding to Si sublimation is too small to quantify a Si:B/N ratio as the amount of Si sublimation is smaller than the typical XPS sensitivity (<1 atomic %).

We now turn to an optical characterisation of the as-grown hBN films. Note, no further plasma cleaning or annealing was performed after the HT-MBE growth. We conducted PL measurements of the series of Si-doped hBN samples and compared it to the grown pristine hBN. Figure 2(a-c) displays a representative $20 \times 20$ $\mu m^2$ PL map of S2-S4, demonstrating that emission spots are spatially distributed over the hBN films. A higher density of isolated single photon emitters (SPEs) in the wide spectral range of 570 - 750 nm was observed in S3 and S4, whereas SPEs were sparsely distributed in sample S2. For comparison, we performed a statistical analysis to monitor emitter densities and corresponding spectral range in samples (S2-S4) as shown in the (Figure S1).

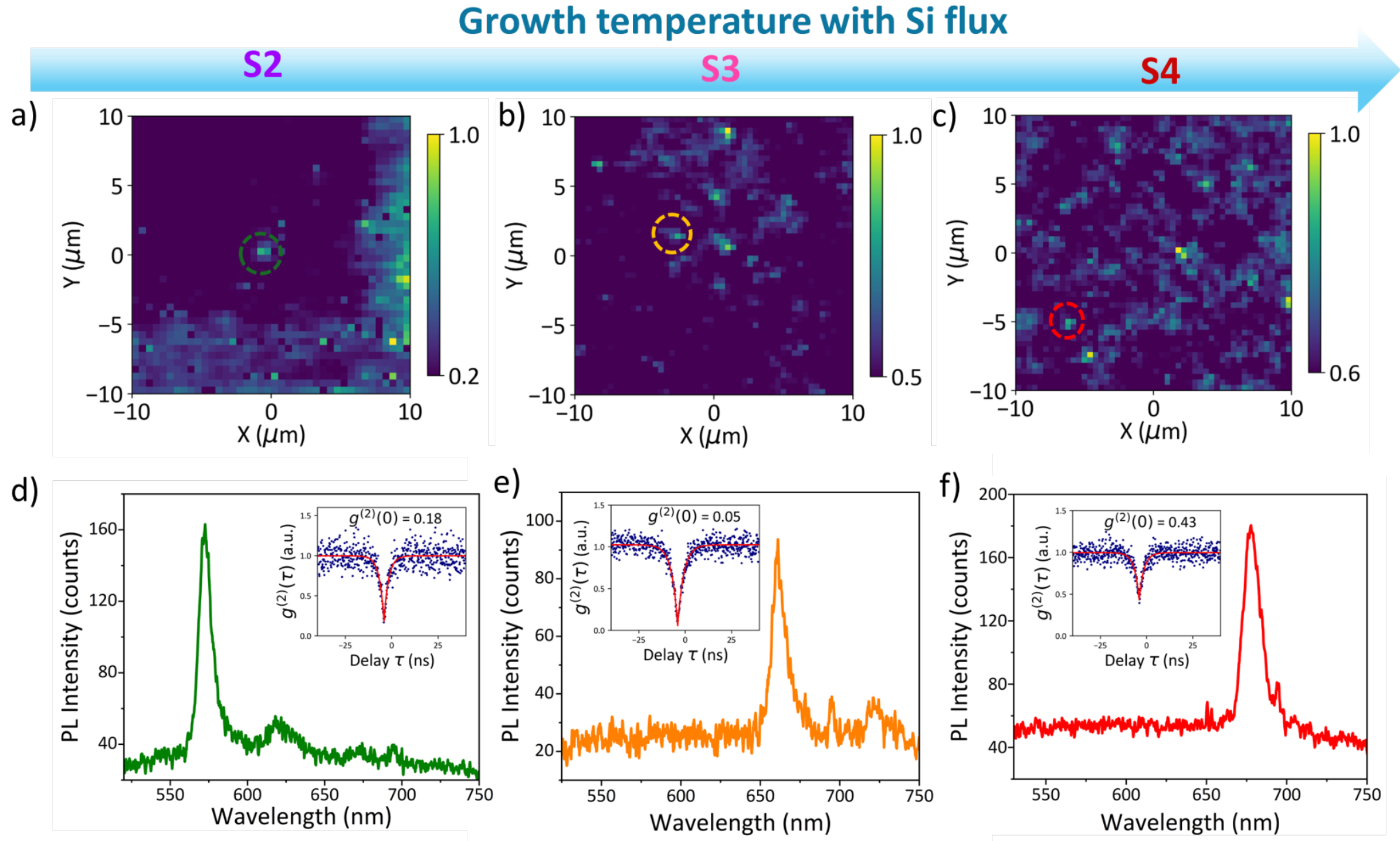

*Figure 2. Single photon emitter (SPE) characterization of as-grown Si-doped hBN samples (S2, S3, and S4) grown at increasing temperatures. a)-c) Confocal PL maps (20 × 20 μm²) of S2, S3, and S4 under 532 nm laser excitation. Isolated bright spots are visible across all samples, with the circled emitters selected for further spectral and photon statistics analysis. d)-f) Representative PL spectra collected from the circled emitters in each sample in the visible wavelength. Insets display the second-order photon autocorrelation function $g^{(2)}(\tau)$ measured via Hanbury Brown and Twiss (HBT) interferometry. The extracted $g^{(2)}(0)$ values of 0.18 (S2), 0.05 (S3), and 0.43 (S4) without background correction.*

Representative spectra from the SPEs are shown in Figure 2(d-f). The SPEs reveal a narrow zero phonon line (ZPL), with some of them showing a pronounced phonon sideband energy around ~ 175 meV. Distinct narrow ZPL wavelengths of 570, 665, and 680 nm corresponding to S2, S3, and S4 are shown as examples. Surprisingly, the observed emitters are relatively narrow with an average FWHM of ~11 nm at room temperature, and they exhibit a clear antibunching dip below 0.5, confirming they are isolated SPEs. We further analyzed polarization dependent measurement of a selected SPE (emits at 728 nm with a FWHM of ~ 10 nm) from S4, which indicates a characteristic of a linearly polarized dipole transition (Figure S2). Typically, such narrow lines in hBN necessitate high temperature annealing, while as grown hBN samples (e.g. by Chemical vapor deposition (CVD) or metalorganic vapour-phase epitaxy (MOVPE)) result in much broader spectral lines.[11,14] Further, we note that no distinct emission was observed in S1 (Figure S3), suggesting that growth temperature plays an important role in defect-formation thermodynamics. To scope out the importance of growth temperature, we also tested the hBN film grown at 1390°C with half of the growth time (labelled as S5) and confirmed S5 sample exhibits similar density of emitters and spectral distribution as S3 and S4.

The temperature-dependent emitter activity and spectral distribution across the sample series provide important information on defect identity. S1 exhibits no SPE activity, which is consistent with a crystalline quality regime with a large density of non-radiative recombination channels. S2 shows low-density emitters with narrow ZPL distribution confined to 570–580 nm, which could be attributed to a lack of thermal energy to activate the formation of more complex defects, vacancy associates, out-of-plane dimers, or charged complexes. As growth temperature increases above 1000 °C, both emitter density and spectral diversity dramatically increase with ZPLs distributed continuously across 550–750 nm. This broadening could be ascribed to the thermally activated formation of various Si-containing configurations, leading to distinct ZPL energies within the observed window. For clarity, Table 1 summarises the studied samples, including their temperature, thickness and SPE activity.

| | **Growth temperature (°C)** | **Time (h)** | **Si flux (A)** | **Thickness (nm)** | **SPE activity** | **Emission (nm)** |
|---|---|---|---|---|---|---|
| **S0 (Ref)** | 1390 | 4 | 0 | 13 | No | - |
| **S1** | 900 | 4 | 30 | 63 | No | - |
| **S2** | 1000 | 4 | 30 | 34 | Yes | 570-580 |
| **S3** | 1200 | 4 | 30 | 16 | Yes | 550-750 |
| **S4** | 1390 | 4 | 30 | 16 | Yes | 550-750 |
| **S5** | 1390 | 2 | 30 | 6 | Yes | 570-750 |

*Table 1. Growth parameters and emitter characteristics of MBE-grown hBN films.*

One of the main challenges in MBE growth of hBN is the undesired incorporation of carbon impurities, which can also act as optically active defect centers.[15,16] We therefore recorded additional PL measurements on S0 sample and confirmed the absence of any bright emitters or sharp ZPL features in the PL map and spectra, suggesting that the MBE-grown hBN matrix itself does not host a significant density of optically active impurities (Figure S4). Additional evidence stems from cathodoluminescence spectroscopy and UV-PL measurement (Figure S5 and S6) that confirm no strong emission at 4.1 eV associated with the in-plane carbon dimer was observed. This UV emission is typical for carbon doped hBN and appears in most hBN crystals that host high density of SPEs. While we can not unambiguously rule out the possibility that the observed SPEs are carbon-related, our results convincingly demonstrate that Si impurities are involved in the emitting complexes.

To comprehensively characterize the ZPL distribution of the observed SPEs, confocal PL mapping and spectra of sample S3 were performed under two different excitation wavelengths (405 nm and 532 nm). In Figure S7, the confocal PL maps reveal spatially localized emitters distributed across the 40 × 40 $\mu m^2$ area under 405 and 532 nm, respectively. Under 405 nm excitation, we observed sharp ZPLs across the visible range including high ZPL energies at ~2.64 eV, while a subset of emitters with ZPLs predominantly distributed between 560 and 650 nm was observed under 532 nm excitation.

As a next step, we tested whether the emitters are indeed incorporated within the hBN film. This is an important measurement, since quantum emitters can be observed in sapphire/alumina or in closely related nitrides.[17-19] As a proof of concept, a polyvinyl acetate (PVA)-assisted dry transfer method was implemented as shown in Figure 4a. The Si-doped hBN films were transferred onto a bare Si/$SiO_2$ substrate.[20] The optical contrast of the transferred hBN film with a lateral scale of hundreds of microns is clearly visible in Figure 4a. The thickness of the transferred film is measured to be ~10 nm by AFM image, as shown in Figure 4b. The thickness is similar to the hBN grown on sapphire, as shown in table 1.

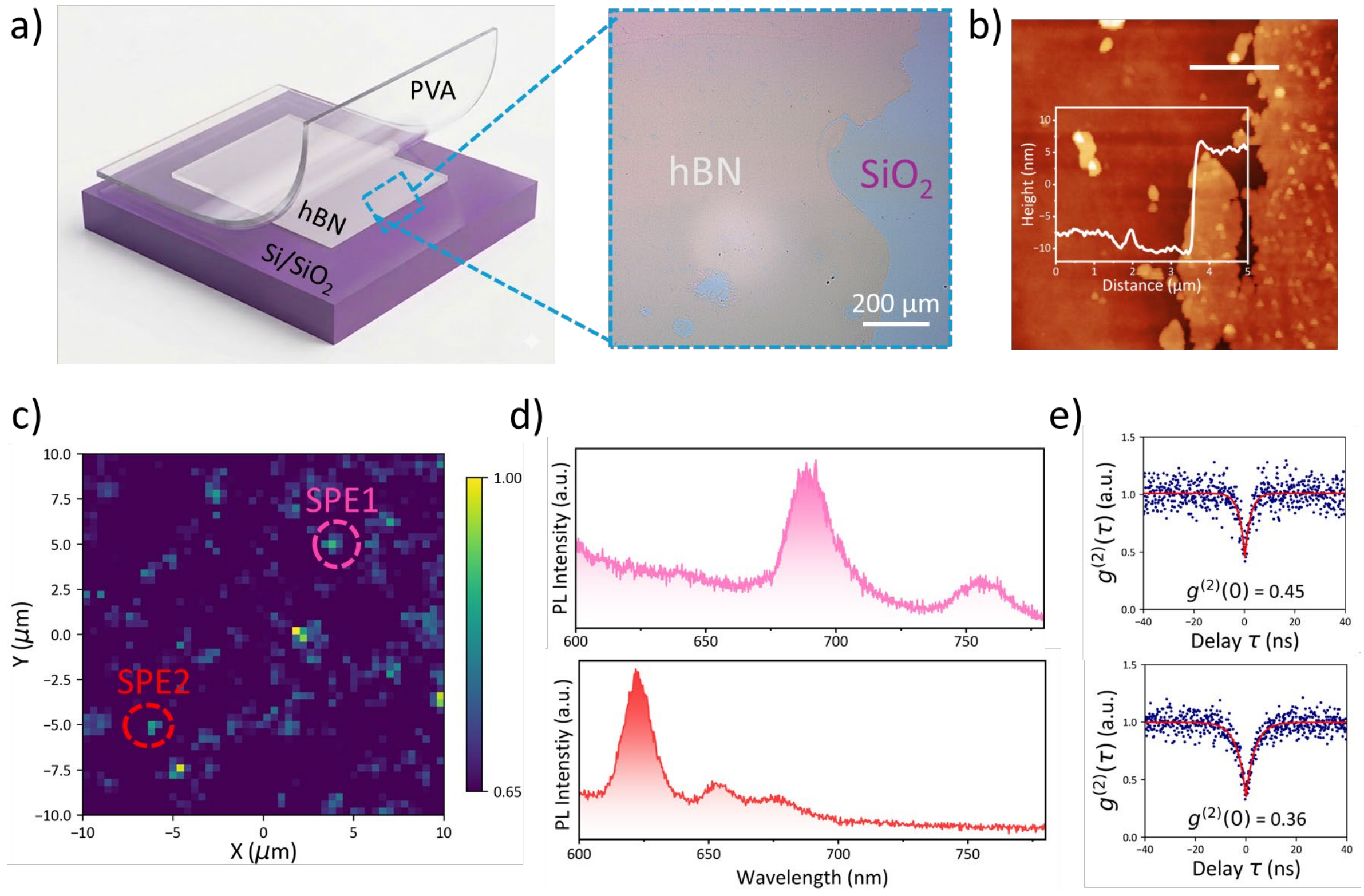

*Figure 3. SPEs characterization of transferred hBN film on Si/SiO₂ substrate. a) Schematic illustration of the PVA-assisted dry transfer of hBN from S4 sample onto a Si/SiO₂ substrate with an optical microscopy image of the transferred flake. b) AFM topography image of the hBN film edge with a thickness profile inset, confirming a step height of ~15 nm. c) Confocal PL map (20 × 20 µm²) acquired under 532 nm excitation. Two isolated SPEs, SPE1 (pink) and SPE2 (red), are highlighted with dashed circles for individual characterization. d) PL spectra of SPE1 (top, ~690 nm) and SPE2 (bottom, ~630 nm), each displaying a sharp ZPL. Second-order photon autocorrelation measurements $g^{(2)}(\tau)$ for SPE1 and SPE2, yielding $g^{(2)}(0)$ values of **0.45** and **0.35**, respectively.*

A confocal PL map of the transferred hBN on $SiO_2$ was carried out at room temperature, and the presence of multiple emitters is confirmed, as shown in figure 4c. Two isolated SPEs are labelled as SPE1 and SPE2 in Figure 4c, and corresponding PL spectra were recorded, showing ZPLs located at 690 and 630 nm, respectively (Figure 4d). To ascertain the quantum nature of SPE1 and SPE2, we recorded the second order correlation function, with $g^2(0)$ of ~ 0.45 and 0.35, respectively. This confirms the emitters originate from hBN films, rather than embedded within the sapphire (Figure S8).

Finally, we discuss the potential chemical configuration of the defects. From a defect-physics perspective, Si incorporation is expected to perturb the local $sp^2$ bonding network more strongly than lighter impurities such as carbon, owing to its larger covalent radius and partial preference for tetrahedral coordination. Density functional theory studies have established that substitutional Si at boron sites ($Si_B$) is energetically more favourable (~2.88 eV) than Si substitution at nitrogen sites (~8.15 eV), producing pronounced out-of-plane lattice relaxation and introducing localized electronic levels within the hBN bandgap.[21-23] The possibility of Si substitution is further corroborated by Annular Dark-Field Scanning Transmission Electron Microscopy (ADF-STEM) analysis. Figure 4a shows a high resolution HAADF-STEM image of a Si-doped hBN with an electron diffraction pattern indexed to the (100) direction, verifying the hexagonal hBN lattice structure. The atomic column along (210) direction within the designated region (marked as dotted square in Figure 4b) exhibit two bright atoms labelled as A and B. The intensity line profile corresponding to A and B along the other atoms in the designated region reveals two times higher peak intensities than the averaged intensity of other atoms. Such significant increase in intensity reflects the presence of Si species occupying the sublattice.

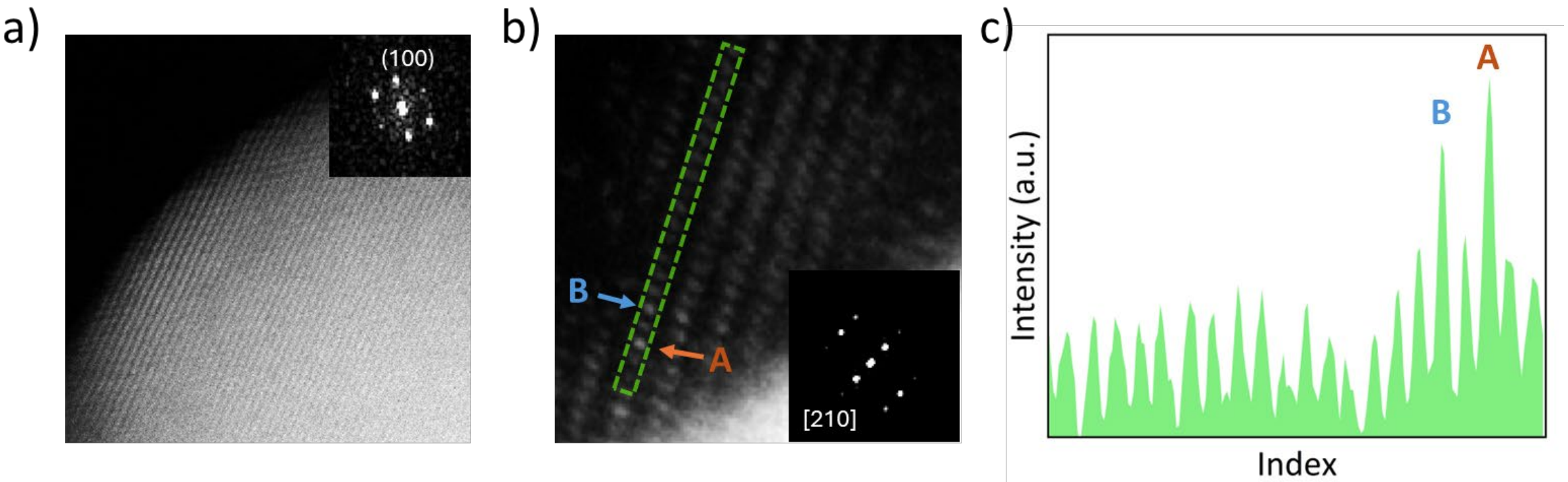

*Figure 4. ADF-STEM image of Si-doped hBN and intensity line scan of atomic column sites. a) High-resolution ADF-STEM image with an inset of diffraction patterns of (100) direction showing the Si-doped hBN flake has the characteristic hexagonal structure. (b) The atomic column resolved ADF-STEM image viewed along [210] direction. Two bright spots labelled as A and B indicate the presence of substitutional Si atoms. c) The intensity line scan within the designated area (dotted square in Figure 4b)) demonstrates that two bright spots, A and B, exhibit ~2 times higher intensity compared to the averaged intensities of other atomic column sites.*

Among the higher-complexity candidates most relevant to the broad ZPL distribution in hBN in S3 and S4 samples, the out-of-plane dimer configurations $Si_NSi_i^-$ and $C_NSi_i^-$ are plausible as they can sustain localized triplet spin manifolds with calculated ZPL spanning approximately 1.5–2.5 eV. As carbon is a common residual impurity in MBE systems, the presence of both C and Si forming planar configurations (e.g. $Si_BC_N$) may also introduce substantial local lattice reconstruction with partially occupied in-gap states. Alternatively, it may arise from DAP recombination since the DAP model has been extensively suggested as the origin for the SPEs in hBN.[24,25] Indeed, as we confirmed the involvement of Si atoms in our hBN film, $Si_B^+$ as donors and native B/N acceptor defects (e.g. $V_B^-$ and $N_BV_N$) in DAP pairs is plausible. Vacancy-associated configurations such as $Si_BV_N$, are less likely to be primary contributors to the narrow ZPLs observed in S3 and S4 since vacancy-related sites in hBN are generally predicted to exhibit large Huang–Rhys factors and broad phonon-coupled emission.[26] All the plausible defect complexes related to Si substitute is summarised in Table S1. At present, however, direct experimental correlation between specific silicon-related crystallographic configurations and individual optical lines remains unresolved. Combined spectroscopy and atomic-resolution microscopy studies are required to distinguish between competing models, as demonstrated by the recent work that isolated the carbon dimer in split interstitial configuration as the 2.8 eV defect.

In conclusion, we have demonstrated an in-situ Si-doped hBN by HT-MBE that hosts localized quantum emitters. The growth temperature was systematically controlled from 900 to 1390 °C at a constant silicon flux, revealing that higher temperatures (>1200 °C) enable a wide distribution of narrow ZPLs across 430-750 nm at room temperature without any further annealing or post-treatment. Moreover, we performed UV-PL and transferred the hBN film onto $Si/SiO_2$ substrate to verify that carbon dimer and surface interaction are unlikely to act as alternative origins. By leveraging ADF-STEM analysis, we proposed silicon-related defect complexes as responsible for the observed SPEs, including DAP recombination with $Si_B^+$ as donor, higher-complexity out-of-plane configurations such as $Si_NSi_i^-$ and $C_NSi_i^-$ as candidate defects governing the ZPL energies. These results give insights into different sources of impurities that offer optically active defects and understanding of impurity-driven defect formation for quantum photonic applications.

**Acknowledgments**

The authors acknowledge Viktor Ivády for the valuable discussions about Si-related defects and Evan Williams for assistance with AFM images. The authors acknowledge financial support from the Australian Research Council (CE200100010, FT220100053, DP250100973, DP260102670) and the Air Force Office of Scientific Research (FA2386-25-1-4044). The authors acknowledge Mr Evan Williams for assistance with AFM measurements. This work at Nottingham was supported by the Engineering and Physical Sciences Research Council UK (Grant Nos. EP/V05323X/1 and EP/W035510/1). This work at Montpellier was supported by the French Agence Nationale de la Recherche through the project BIRD (ANR-24-CE08-3820).

**Conflicts of Interest**

The authors declare no conflicts of interest.

**Data Availability Statement**

The data that support the findings of this study are available from the corresponding author upon reasonable request.